\documentclass[prb,twocolumn,amsmath,amssymb]{revtex4}
\usepackage{graphicx}
\usepackage{dcolumn}
\usepackage{bm}

\newcommand{\lpcmo}{(La$_{1-y}$Pr$_{y}$)$_{0.7}$Ca$_{0.3}$MnO$_3$}
\newcommand{\lpcmoa}{(La$_{0.5}$Pr$_{0.5}$)$_{0.7}$Ca$_{0.3}$MnO$_3$}
\newcommand{\lpcmob}{(La$_{0.25}$Pr$_{0.75}$)$_{0.7}$Ca$_{0.3}$MnO$_3$}
\newcommand{\lcmo}{La$_{1-x}$Ca$_{x}$MnO$_3$}
\newcommand{\tc}{$T_C$}
\newcommand{\tg}{$t_{2g}$}

\newcommand{\cm}{cm$^{-1}$}
\newcommand{\sdc}{$\sigma_{DC}$}
\newcommand{\sig}{$\sigma_1(\omega)$}
\newcommand{\eps}{$\epsilon_1(\omega)$}

\begin{document}

\title{Oxygen isotope effect and phase separation in the optical conductivity of
\lpcmoa\ thin films}

\author{F. P. Mena$^{1}$, A. B. Kuzmenko$^{2}$, A. Hadipour$^{1}$, J.L.M.
            van Mechelen$^{2}$, D. van der Marel$^{2}$ and N. A. Babushkina$^{3}$}
\affiliation{   $^{1}$ Material Science Center, University of
                Groningen, 9747 AG Groningen, The Netherlands.\\
                $^{2}$ D\'{e}partement de Physique de la Mati\`{e}re
                Condens\'{e}e, Universit\'{e} de Gen\`{e}ve,
                 quai Ernest-Ansermet CH - 1211 Gen\`{e}ve,
                 Switzerland.\\
                $^{3}$ RRC Kurchatov Institute, 123182 Moscow, Russia.}
\date{\today}

\begin{abstract}
The optical conductivities of films of \lpcmoa\ with different
oxygen isotopes ($^{16}$O and $^{18}$O) have been determined in
the spectral range from 0.3 to 4.3 eV using a combination of
transmission in the mid-infrared and ellipsometry from the
near-infrared to ultra-violet regions. We have found that the
isotope exchange strongly affects the optical response in the
ferromagnetic phase in a broad frequency range, in contrast to the
almost isotope-independent optical conductivity above \tc. The
substitution by $^{18}$O strongly suppresses the Drude response
and a mid-infrared peak while enhancing the conductivity peak at
1.5 eV. A qualitative explanation can be given in terms of the
phase separation present in these materials. Moreover, the optical
response is similar to the one extracted from measurements in
polished samples and other thin films, which signals to the
importance of internal strain.
\end{abstract}

\maketitle

\section{Introduction}

\begin{figure}
  \centerline{\includegraphics[width=86mm]{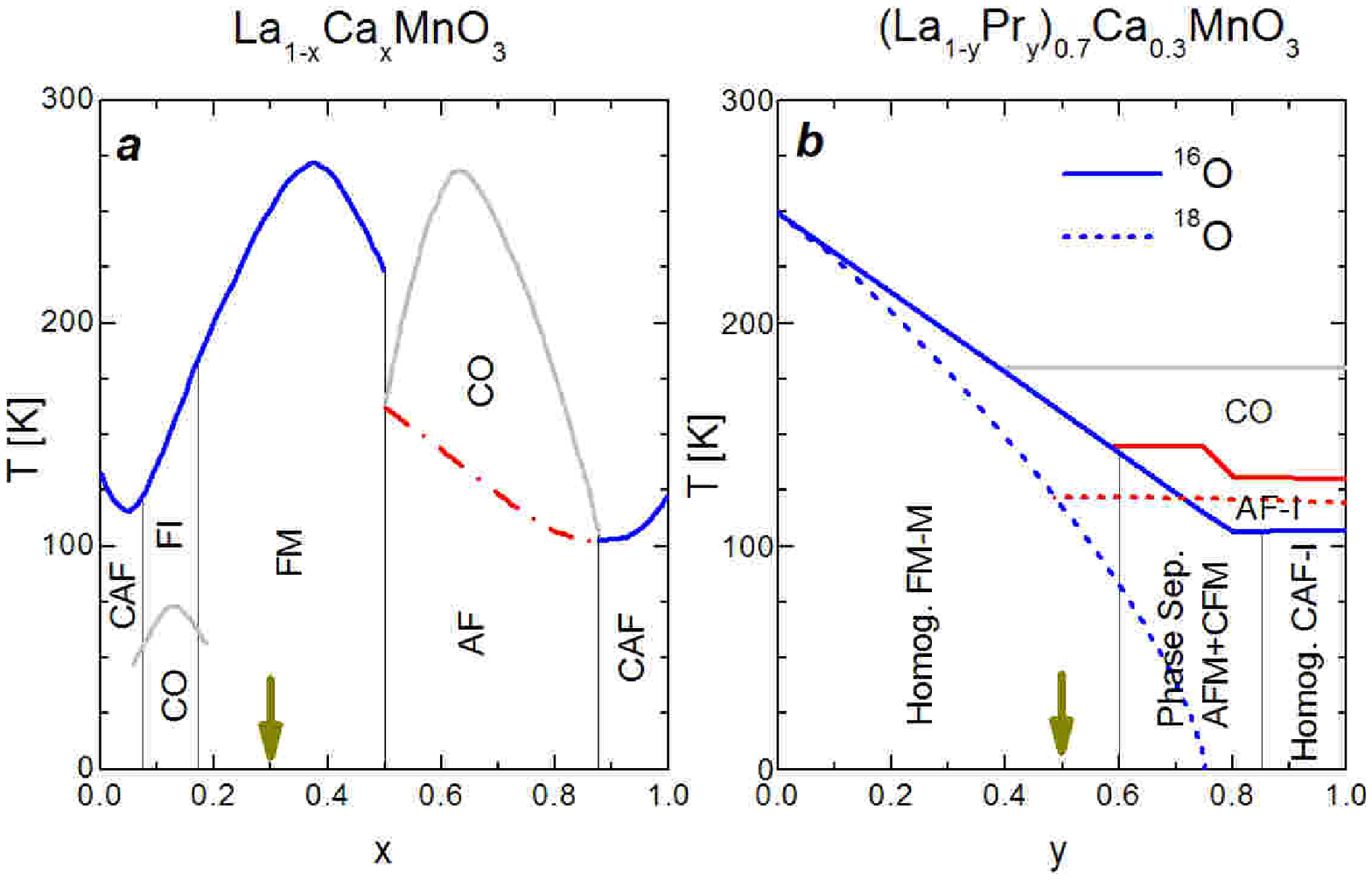}}
  \caption{
  Phase diagram of {\it (a)} \lcmo\ (from Ref.
  \onlinecite{millis-review}), and {\it (b)} \lpcmo\ (from Ref. \onlinecite{balagurov-prb64}
  for $^{16}$O and as suggested in Ref. \onlinecite{babushkina-jpcm11} for
  $^{18}$O). The richness of these phase diagrams is evident. They
  comprise the following states: canted antiferromagnet (CAF),
  charge order (CO), ferromagnetic insulator (FI),
  ferromagnet(FM), and antiferromagnet (AF). The arrow in panel {\it b} indicates
  the composition used in this work. However, it has to be taken
  into account that due to internal strains, the film actually
  lies closer to the region of phase separation. \cite{babushkina-epjb19}
  }\label{mang-phase-fig}
\end{figure}

Few compounds manifest a so ample variety of phases as the manganite
perovskites when conditions such as magnetic field, temperature, or
doping are changed. \lcmo, a particular well studied family of
manganites, is an example of this as can be seen from its phase
diagram given in Fig.~\ref{mang-phase-fig}a.\cite{millis-review} It
is evident that the particular ground state of the manganites
depends on the ratio of Mn$^{3+}$ to Mn$^{4+}$ given by the hole
doping $x$. At a doping between about 0.17 and 0.5 the
low-temperature phase is metallic and ferromagnetic, which is
commonly attributed to the double exchange mechanism. \cite{zener}

The ground state also depends on specific lattice characteristics:
Mn-O and cation-O bond lengths, and cation-O-cation bond angles.
These bond characteristics depend, in turn, on the average cation
radius, $\langle r_A \rangle$.\cite{hwang} At a given hole doping,
the direct influence of $\langle r_A \rangle$ can be studied by
replacing the rare earth with another. However, this replacement
can also induce disorder if $\langle r_A \rangle$ exceeds certain
critical value.\cite{rodriguez} \lpcmo\ is particularly suitable
to study the effect of the average cation radius as it always
remains below the critical value.\cite{balagurov-prb64} Neutron
diffraction experiments in this system\cite{balagurov-prb64} have
revealed the phase diagram shown in Fig.~\ref{mang-phase-fig}b
(solid lines). At low temperatures, as the Pr concentration
increases, the system goes from a homogeneous ferromagnetic metal
to a homogeneous canted antiferromagnetic insulator. Between these
two extremes, the neutron diffraction data\cite{balagurov-prb64}
suggests the presence in the system of two different phases, one
is antiferromagnetic while the other is canted ferromagnetic.

Phase separation seems, indeed, to be a common feature of strongly
correlated systems (see Ref. \onlinecite{khomskii} and references
therein). For example, the double-exchange model has a natural
tendency towards phase separation at low doping.\cite{khomskii,
moreo, kagan} The regions formed in this way are one of an undoped
antiferromagnet and one of higher electron (or hole) concentration
which is ferromagnetic (or strongly canted) and
metallic.\cite{khomskii} However, the long-range Coulomb forces
compete against phase separation forcing the phase domains to be
rather small.\cite{khomskii, moreo} It is important to mention
that there can also be phase separation between electron rich
regions and charge ordered regions. \cite{khomskii} Several
experimental techniques yield a strong evidence for phase
separation in the manganites (see Refs.~\onlinecite{khomskii} and
\onlinecite{moreo} for a review, and
Ref.~\onlinecite{phase-sep-films} for experiments specific to thin
films).

Another striking phenomenon observed for certain compositions is
the strong oxygen isotope effect.
\cite{zhao-nature,babushkina-nature} For instance, the effect of
oxygen isotope substitution ($^{16}$O $\rightarrow$ $^{18}$O) has
been studied in various members of the \lpcmo\ family.
\cite{babushkina-jpcm11, babushkina-nature} In the paramagnetic
state neither the lattice parameters, \cite{balagurov-prb60} nor
the DC conductivities \cite{babushkina-jpcm11, babushkina-nature}
are noticeably modified by the isotope exchange. At low
temperatures, however, the situation is drastically different.
Samples containing heavier $^{18}$O atoms present lower \tc's and
less metallic conductivity. In particular, for $y=0.75$, the
isotope substitution does even induce a metal-insulator
transition.\cite{babushkina-nature} Thus the lattice dynamics
clearly plays an important role in the charge transport and spin
ordering in manganites.

Several mechanisms have been formulated to explain the isotope
effect. Alexandrov and Bratkovsky described the isotope dependence
in terms of polaron-bipolaron dynamics.\cite{alexandrov}
Alternatively, it has been attributed to a slight oxygen-mass
dependence of the lattice Debye-Waller factor and, therefore, the
hopping parameter $t$ and bandwidth $W$, which is close to the
critical value.\cite{babushkina-jap83} In order to be so sensitive
to the oxygen mass, the system has to be at the verge of the
metal-insulator transition. This sensitiveness can also be related
to phase separation.\cite{khomskii} As seen in
Fig.~\ref{mang-phase-fig}b, \lpcmob\ (with $^{16}$O) seems to be
in a phase-separated state. Then, the small changes produced by
the isotope substitution would favor one phase over the other
producing the metal-insulator transition.

The isotope effect has also been studied in thin films of \lpcmoa\
grown in either LaAlO$_3$ (LAO) and SrTiO$_3$ (STO).
\cite{babushkina-prb59, babushkina-epjb19} For the films grown in
LAO, the isotope substitution produces a metal-insulator
transition similar to the one seen in ceramic samples of \lpcmob.
\cite{babushkina-epjb19, babushkina-prb59} Although for films
grown in STO no metal-insulator transition was observed, the
samples with $^{18}$O exhibit a much lower \tc\ and the value of
the DC conductivity (\sdc) at low temperatures is one order of
magnitude lower than those containing $^{16}$O. The sensitivity to
the substrate material arises from the different strains they
support due to mismatches between the lattice parameters of the
film and the substrate. \cite{babushkina-epjb19} When grown in
LAO, the films are contracted in plane but stretched
perpendicularly. The opposite is true for films grown in STO.
These deformations are reflected in the magnitude of the Mn-O-Mn
bond angles. Compared to the ceramic samples, this angle is
increased for films in STO and decreased for LAO. This structural
difference puts the films grown in LAO at an angle closer to the
critical value corresponding to localization of carriers (remember
that the conduction bandwidth depends on the cosine of the Mn-O-Mn
angle). \cite{babushkina-epjb19}

Optical spectroscopy, which probes charge dynamics, has played an
important role in studying the physics driving the behavior of
manganites. \cite{millis-sw-cmr} It has been found
\cite{okimoto,quijada} that the metal-insulator transition is
accompanied by a large redistribution of optical spectral weight,
mainly below $\sim$ 4 eV. It indicates the energy scale of the
most important electronic interactions making up the phase
diagram, such as the on-site Hunds's rule interaction or polaron
activation energy. The value of the electronic kinetic energy,
derived from the integrated optical conductivity, has been a key
ingredient of the quantitative verification of the double-exchange
scenario in manganites.\cite{millis-sw-cmr,michaelis}

Optical conductivity has also offered some indications of phase
separation occurring at some specific concentrations. \cite{moreo}
For example, let us consider La$_{7/8}$Sr$_{1/8}$MnO$_3$
\cite{jung-lsmo} which remains insulating even below \tc. Its
optical conductivity does not show any Drude contribution but a
mid-infrared (MIR) peak appeared at around 0.4 eV below \tc.
\cite{jung-lsmo} This peak was assigned to a small polaron
absorption and it was noticed that the temperature dependence of
its spectral weight resembles a percolation-type transition.
\cite{jung-lsmo} In fact, Moskvin et al. \cite{moskvin-effmedium}
have been able to describe the temperature dependence of the
optical conductivity in this compound by using an effective medium
approximation assuming that metallic spherical regions are
embedded into an insulating matrix (LaMnO$_3$). In this
approximation the MIR peak is a geometric resonance whose position
is mainly determined by the shape of the metallic regions.
\cite{moskvin-effmedium}

Such MIR peak has also been seen in the FM metallic part of the
phase diagram. \cite{kim-lcmo, kim-lpcmo, jung-lcmo} It is already
visible at high temperatures but below \tc\ its intensity
increases being also accompanied by a narrow Drude peak. This
behaviour suggested the change from a small to a large polaron.
\cite{kim-lcmo} However, in contrast to the experiments performed
on polished samples, experiments on cleaved single crystals
\cite{takenaka-cleaved} show, the development of only a broad
zero-centered peak. The difference in the results emphasize the
sensitivity of the manganites to static imperfections and/or
structural strain. \cite{takenaka-cleaved}

Optical experiments in thin films have also shown the presence of
a narrow Drude-like mode and a MIR peak in FM-metallic samples
(see, for example, Refs. \onlinecite{quijada} and
\onlinecite{hartinger}). These MIR peaks have also been assigned
to polaronic absorptions. Particularly, it was possible to
distinguish between small and large polarons in
La$_{2/3}$Ca$_{1/3}$MnO$_{3}$ and La$_{2/3}$Sr$_{1/3}$MnO$_{3}$,
respectively. \cite{hartinger} In contrast to the work of Kim et
al.\cite{kim-lcmo} (discussed above), Hartinger and collaborators
made the assignment by fitting the optical conductivity to the
corresponding theoretical expressions. \cite{hartinger} Evidence
of phase segregation has also been inferred from measurements of
the absorption coefficient in manganite thin films (see Refs.
\onlinecite{moskvin-jt, moskvin-lcmo, loshkareva-isotope} and
references therein).

In the present article we present for the first time, to our
knowledge, the optical conductivity of manganites containing two
different oxygen isotopes. Specifically, we studied films of
\lpcmo\ ($y = 0.5$) grown on SrTiO$_3$. We will see that the
isotope substitution produces changes in the optical conductivity
at energies much larger than the phonon region. Both samples show
a strong MIR infrared peak whose intensity decreases when
temperature is increased. The spectral weight lost by this peak is
mainly transferred to a peak lying at around 1.5 eV. By a detailed
study of their temperature dependence we will argue that this
behavior is consistent with phase separation.

\section{Experiment}\label{mang-exp}

\subsection{Sample Preparation}\label{mang-exp+sample}

Thin films of \lpcmoa\ were grown on SrTiO$_3$ using the aerosol
MOCVD technique (for the details of the preparation and
characterization see Ref. \onlinecite{babushkina-epjb19}). The
films had a nominal thickness of 60 nm which was extracted more
accurately from the optical measurements. For the isotope exchange
two strips of ~$1 \times 8$ mm$^2$ were annealed simultaneously in
different atmospheres. One of them was heated in a $^{16}$O$_2$
atmosphere, while the other was heated in an oxygen atmosphere
containing 85\% of $^{18}$O$_2$. Hereafter, these samples will be
referred to as S16 and S18, respectively. The XRD analysis shows
that the films are highly strained. \cite{babushkina-epjb19} As
mentioned in the introduction, the STO lattice constants are
larger than the film, which produces an in-plane expansion of the
perovskite cube. In contrast, perpendicularly to the plane, the
film is contracted. In this conditions, a buckling of the MnO$_6$
octahedrons is expected. \cite{babushkina-epjb19}

\subsection{Optical Experiments}\label{mang-optexp}

\subsubsection{Transmission}

\begin{figure}
  \centerline{\includegraphics[width=86mm]{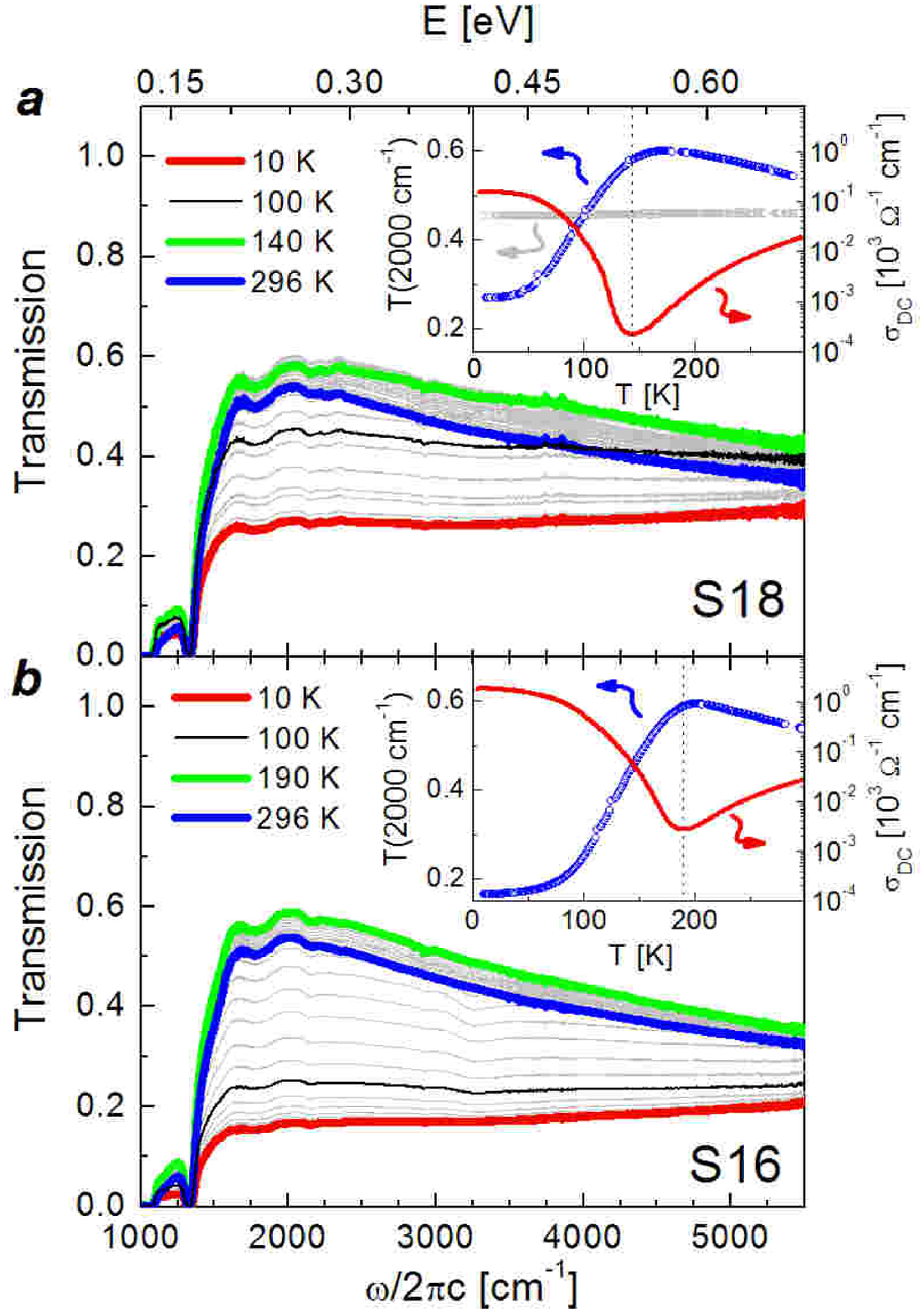}}
  \caption{
  Transmission of  two \lpcmoa\ films with $^{16}$O (S16) and $^{18}$O (S18)
  (thickness $\sim$ 60 nm) grown on STO substrates.
  The spectra presented in the main panels correspond
  to temperatures between 10 and 296 K every 10 K .
  {\it Insets:} Temperature dependence of the transmission at 2000 \cm\
  and the DC conductivity, both measured in the heating mode with no
  magnetic field. The vertical dotted line indicates the minimum in
  \sdc\ which is associated with the entrance to the
  ferromagnetic state (we will refer to it as \tc). The inset of the upper panel also shows the
  temperature dependence of the transmission (shifted down by 30\%)
  of a STO substrate. It is clear that the temperature effects
  are solely due to the changes in the films.
  }\label{mang-trans-fig}
\end{figure}

In the frequency region 1000 - 5000 \cm\ the optical transmission
was measured using a Bruker 113v FT-IR spectrometer. Below this
frequency the STO substrate is not transparent. The transmission was
calibrated at room temperature against an aperture of the same size.
For the temperature dependence of the transmitted intensity a
home-built cryostat was used, the special construction of which
guarantees the stable and temperature independent optical alignment
of the sample. The measured intensities where then normalized to the
room-temperature transmission. The results of these measurements are
summarized in Fig.~\ref{mang-trans-fig}. The main panels show the
absolute transmission measured while the temperature was increased.
The insets show the temperature dependence of the transmissions at
2000 \cm\ and \sdc\ measured also in heating mode. \sdc\ in both
films shows hysteresis, being rather large in the S18
sample\cite{babushkina-epjb19}, which is probably an indication of
the phase separation (see below).

\subsubsection{Substrate}

\begin{figure}
  \centerline{\includegraphics[width=86mm,clip=true]{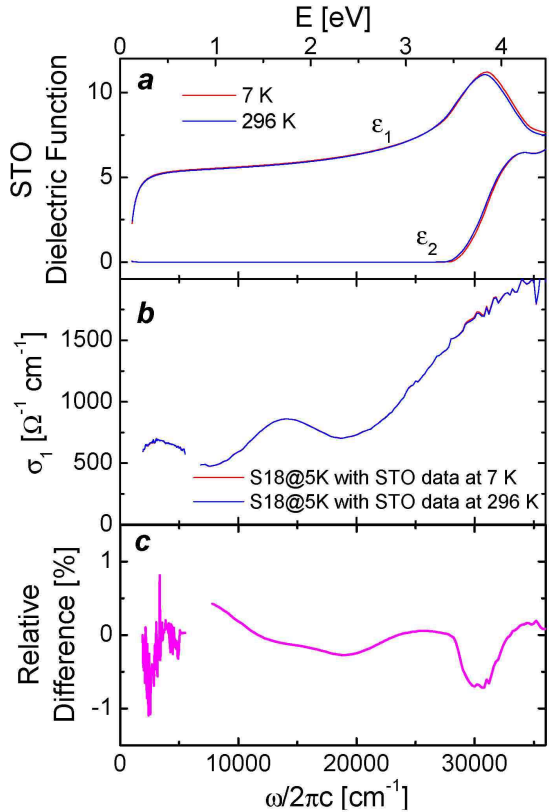}}
  \caption{
    {\it (a)} Measured complex dielectric function of STO at 296 and 7 K.
    {\it (b)} Optical conductivity of the S18 filmobtained from the raw data
    at 5 K and using the STO dielectric function at 296 (blue) and 7 K (red).
    {\it (c)} Relative difference, expressed in percentage, between the two obtained
    conductivities. This result indicates the negligible effect of the
    temperature variation of the substrate.
  }\label{mang-stodiff-fig}
\end{figure}

The optical properties of the STO substrate were determined below
6000 \cm\ using the combination of reflectivity and transmission.
The complex dielectric function was then determined by numerically
inverting the corresponding Fresnel equations. \cite{heavens}
Moreover, the temperature dependencies of transmission and
reflectivity were measured. However, the dependence is much smaller
than the changes in the films. As an example, the transmission of
STO at 2000 \cm\ is shown in the inset of Fig.~\ref{mang-trans-fig}.
Above 6000 \cm, we have used ellipsometry in combination with
transmission in order to obtain the dielectric function. The results
at room temperature and 7 K are shown in the top panel of
Fig.~\ref{mang-stodiff-fig}.

\subsubsection{Ellipsometry}

Ellipsometry of the samples was performed in the range 6000 to
36000 \cm. We used a commercial (Woollam VASE32) ellipsometric
spectrometer in combination with a home-made ultra high vacuum
cryostat working at a pressure of about $10^{-9}$ mbar. At this
pressure, the effects of ice growth were not observable during the
experiment.

The ellipsometry experiment was carried out in a grazing
reflectivity configuration at an angle of incidence of 80$^\circ$.
The necessary alignments were only performed at room temperature
since the special construction of the cryostat allows only a small
displacement of the cold-finger ($\sim0.1$ mm) in the whole
temperature range. This displacement is much smaller than the size
of the sample ($\sim 5 \times 2$ mm$^2$). Moreover, the quartz
windows used in the cryostat had a very small depolarizing effect
of less than 1$^\circ$ for both $\Psi$ and $\Delta$ (see below) as
it was checked separately. This small depolarization is achieved
by an special mounting that minimizes mechanical strains.

\begin{figure}
  \centerline{\includegraphics[width=86mm]{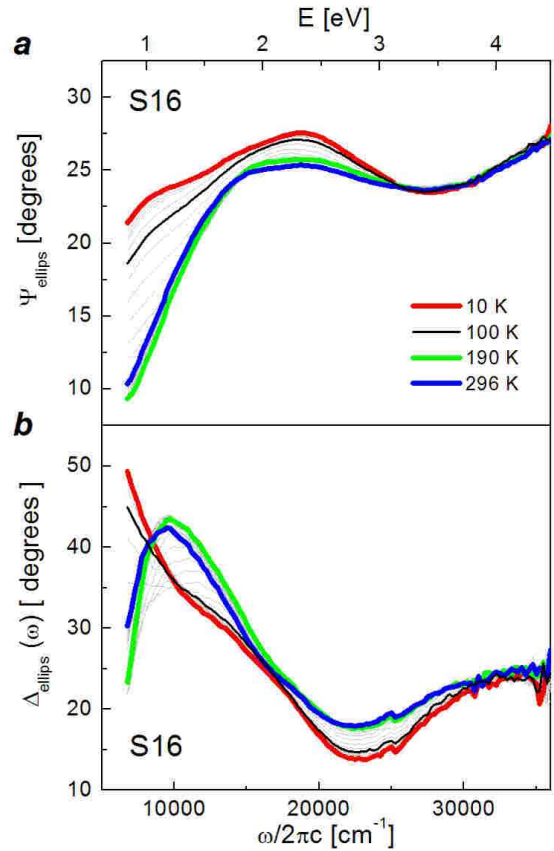}}
  \caption{
    Parameters $\Psi$ and $\Delta$ for S16 as obtained directly
    from ellipsometry at an angle of incidence of 80$^\circ$ between 10 and 296 K every 10 K.
  }\label{mang-ellips-fig}
\end{figure}

\begin{figure}
  \centerline{\includegraphics[width=86mm,clip=true]{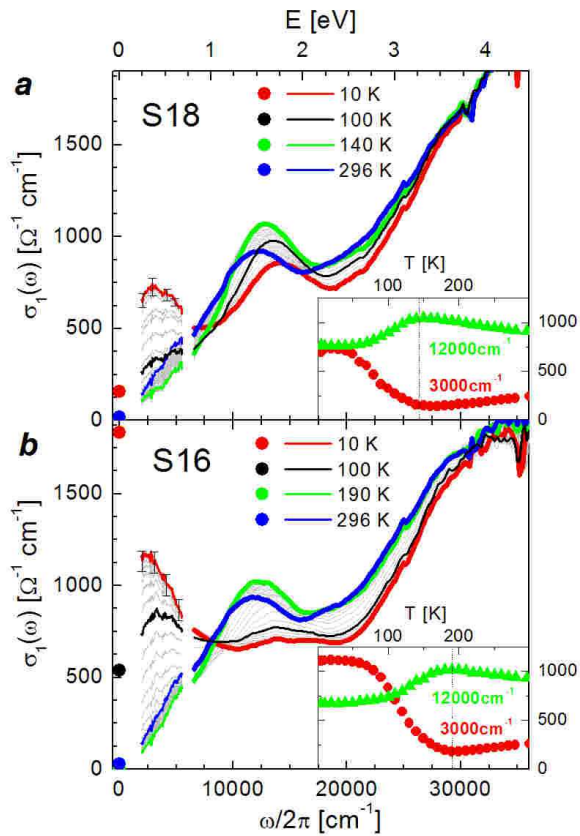}}
  \caption{
    Optical conductivity of the S16 and S18 samples
    between 10 and 296 K every 10 K. {\it Insets:} Temperature dependencies of
    the optical conductivity at 3000 and 12000 \cm.
  }\label{mang-s1-fig}
\end{figure}

The outcome of the ellipsometry experiment are the ellipsometric
parameters $\Psi$ and $\Delta$ that define the ratio between the
Fresnel reflection coefficients for the $s$- and $p$-polarized
light, $\rho \equiv \frac{r_p}{r_s} \equiv \tan(\Psi){\mbox e}^{i
\Delta}$. As an example of the results, we show in
Fig.~\ref{mang-ellips-fig} the corresponding ellipsometric
parameters for S16. The ratio $\rho$, obviously, depends on the
dielectric functions of both, substrate and film. Since we know the
optical properties of the substrate, the complex dielectric function
of the film, $\epsilon(\omega) = \epsilon_1(\omega) + i
(4\pi/\omega)\sigma_1(\omega)$, was obtained by inverting
numerically the analytical expression corresponding to $\rho$ for a
two-layer system. \cite{heavens} At different temperatures we used
only the room temperature data for the substrate and also assumed a
semi-infinite substrate. The real part of the optical conductivity,
\sig, obtained from this inversion is shown in
Fig.~\ref{mang-s1-fig}.

\subsubsection{Optical Conductivity in the MIR Range}

Below 6000 \cm, the complex dielectric function of the film was
obtained using a similar approach to the one used in our previous
experiments. \cite{mena-ms} As a first step we performed a
simultaneous fit to the DC conductivity, transmission of the
film+substrate system (via the Fresnel equations), and the
dielectric function of the film at higher frequencies. The fit
used a model dielectric function which is the combination of one
Drude and a limited number of Lorentz oscillators:

\begin{equation}\label{mang-DL-eq}
  \epsilon (\omega) = \epsilon_\infty -
  \frac{\omega_p^2}{\omega(\omega+i\Gamma)}
  + \sum_j \frac{\omega_{p,j}^2}{(\omega_{o,j}^2 - \omega^2)- i \omega \Gamma_j }
\end{equation}

\noindent where $\omega_o$, $\omega_p$ (plasma frequency), and
$\Gamma$ (scattering rate) give respectively the position, strength,
and width of the oscillator. $\epsilon_{\infty}$ is the
high-frequency dielectric constant. This Drude-Lorentz (DL) fit sets
extrapolations below and above the measured range. Finally, a
Kramers-Kronig constrained variational fitting of spectra is used
where every detail of the measured data is reproduced by introducing
an arbitrary number of oscillators on top of the previous fit
\cite{reffit-manual}.

The real part of the optical conductivities obtained in this way are
plotted in Fig.~\ref{mang-s1-fig}. As in the usual KK
transformation, the obtained optical conductivity depends on the
extrapolations. Since we have used the dielectric function obtained
from ellipsometry at a rather broad energy range, the most
influential are the low frequency extrapolations. To estimate the
uncertainties coming from this factor, we have repeated the
described procedure using different extrapolations (all of them
congruent with the DC conductivity). The associated error bars are
small for \sig\ though rather large for the low-frequency \eps. This
is not a surprise since the transmission is mostly determined by the
absorptive part of the optical conductivity, \sig. Moreover, we have
repeated the same calculations assuming an error as large as 5\% in
the measured transmission. The total uncertainty related with both
sources of errors is indicated by the error bars of
Fig.~\ref{mang-s1-fig}.

The last point we would like to comment in this section is the
influence of the temperature-dependent variation of the substrate.
This is indeed important as it turns out that the penetration depth
of the films is of the order of the film thickness. Therefore, we
have made a complete analysis of such effect in our data. With the
raw data of the substrate+film system at 5K, we repeated the
calculations described above but using the dielectric function of
STO at low temperatures. The result of this analysis is presented in
the middle and bottom panels of Fig.~\ref{mang-stodiff-fig}, which
shows a variation smaller than 1\% of the film optical conductivity.
Consequently we may exclude any spurious effect on the dielectric
function displayed in Fig.~\ref{mang-s1-fig} due to the temperature
dependence of the substrate.

\section{Results and Discussion}

\subsection{Optical Conductivity and Assignment of Peaks}

\begin{figure}
  \centerline{\includegraphics[width=86mm,clip=true]{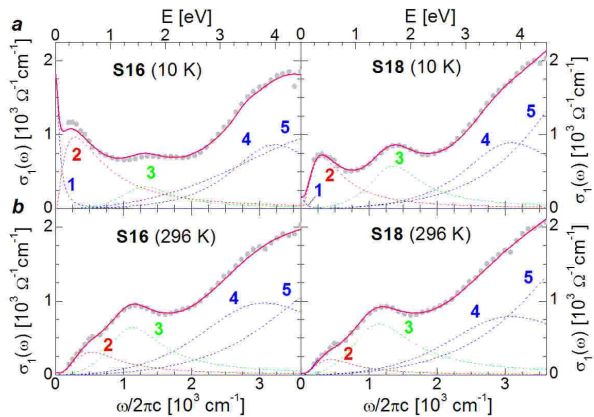}}
  \caption{
    Results of the Drude-Lorentz fit. Optical conductivity (gray circles),
    fit (thin solid line) and individual oscillators
    (dashed lines) at {\it (a)} 10 K, and {\it (b)} room temperature.
  }\label{mang-osc-fig}
\end{figure}

The obtained optical conductivities are shown in
Figures~\ref{mang-s1-fig} and \ref{mang-s1-fig1}. The most important
observation is that, while above \tc\ the samples with $^{16}$O and
$^{18}$O have almost identical \sig\ (Fig.~\ref{mang-s1-fig1}a), at
low temperatures significant differences appear
(Fig.~\ref{mang-s1-fig1}b). Nevertheless, even at low temperatures,
the conductivities of the two isotope-substituted samples
demonstrate qualitatively the same set of peaks
(Fig.~\ref{mang-osc-fig}). This allows us to discuss the provenance
of the main spectral features for the two samples.

In agreement with previous measurements (see, for example,
Ref.~\onlinecite{quijada}), we observe a significant transfer of
spectral weight from the high-frequency region (1 - 4 eV) to low
frequencies as the system goes from the paramagnetic to FM state
(see the opposite trends in the temperature dependence of \sig\ at
3000 \cm\ and 12000 \cm, shown in the insets of
Fig.~\ref{mang-s1-fig}). Also notice that at 3000 \cm\ the optical
conductivity increase below \tc\ is larger in the more metallic
S16 sample. Other point to mention is that the frequency
dependence of the optical conductivity, at different temperatures,
is similar to that obtained from reflectivity measurements on
polished samples.
\cite{jung-lsmo,kim-lcmo,kim-lpcmo,jung-lcmo,okimoto} In
particular, both samples exhibit an MIR peak which seems to be
accompanied by a narrow zero-frequency mode in the S16 sample.
Note that Takenaka et al.
\cite{takenaka-cleaved,takenaka-L825S175MO,takenaka-LSMO} claimed
that non-polished (cleaved) single crystals do not demonstrate any
separation of the Drude and MIR peaks.

The DL oscillator fit described in Section~\ref{mang-optexp}
allowed us to separate the different contributions to the optical
conductivity and their temperature dependencies. At all
temperatures and in both samples, we used the same number of
oscillators. For example, at 10 and 296 K, the oscillators used in
the fit can be seen in Fig.~\ref{mang-osc-fig} while the fit
parameters at 10 K are given in Table~\ref{mang-osc-tab}. The
limited spectral range of our measurements did not allow us to
extract the exact shape of the narrow Drude peak. Therefore we had
to keep the width of the Drude peak constant at a small value and
adjust its strength according to the value of \sdc. In our case
the width of the Drude peak is largely an {\it ad hoc} assignment
but it is in agreement with what has been seen in polished samples
\cite{kim-lcmo} and thin films \cite{hartinger}. Nevertheless, it
does not affect the main conclusions presented below. We have also
fixed the position of the highest-frequency oscillator, which
falls outside our spectroscopic window.

\begin{table*}[t]
  \caption{Parameters, in \cm\ (eV), corresponding to the oscillators
  used to describe the optical conductivity of the two films at 10 K.}\label{mang-osc-tab}
  \begin{ruledtabular}
    \begin{tabular}{c|cc||ccc|ccc}
      Sample & $\omega_{p,1}$ & $\Gamma_1$ & $\omega_{o,2}$ & $\omega_{p,2}$ & $\Gamma_2$ & $\omega_{o,3}$ & $\omega_{p,3}$ & $\Gamma_3$ \\
      \hline
      S16 & 8320 & 632 & 2853 (0.35) & 21154 & 7725 & 13155 (1.63) & 12872 & 9300 \\
      S18 & 2449 & 632 & 3054 (0.38) & 15999 & 6132 & 13559 (1.68) & 18091 & 9426 \\
    \end{tabular}
  \end{ruledtabular}
    \vspace{3mm}

  \begin{ruledtabular}
    \begin{tabular}{c|ccc|ccc}
      Sample & $\omega_{o,4}$ & $\omega_{p,4}$ & $\Gamma_4$ & $\omega_{o,5}$ & $\omega_{p,5}$ & $\Gamma_5$ \\
      \hline
      S16 & 32236 (3.99) & 29631 & 16843 & 43600 (5.41) & 55234 & 47633 \\
      S18 & 30930 (3.84) & 31347 & 18300 & 43600 (5.41) & 53329 & 24540 \\
    \end{tabular}
  \end{ruledtabular}

\end{table*}

\begin{figure}
  \centerline{\includegraphics[width=86mm,clip=true]{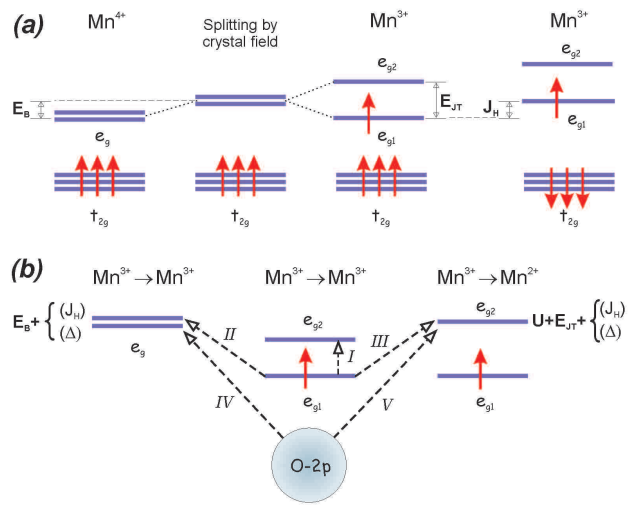}}
  \caption{
    Energy diagram of the manganites (adapted from
    Refs.~\onlinecite{millis-review} and \onlinecite{quijada}). {\it
    (a)} The Mn-d levels, due to the crystal field, are split in the so-called $t_{2g}$ and $e_g$
    levels. If the $e_g$ level is occupied, this state is further split by
    $E_{JT}$, due to the Jahn-Teller distortions of the surrounding O atoms.
    If the $e_g$ level is empty, its original energy
    can be shifted by $E_B$ due to a "breathing" distortion
    which couples to changes in the $e_g$ occupation density.
    Furthermore, the $e_g$ level can correspond to spin states
    either parallel or antiparallel to the total spin of the $t_{2g}$ electrons. In the latter
    case, the energy of $e_g$ is higher by an energy given by
    the Hund's coupling, $J_H$. We have plotted
    this situation for the case of an occupied $e_g$ level.
    {\it (b)} Possible transitions occurring in the manganites.
    The initial states (a single-occupied $e_g$ level
    and a filled O-2p band) are depicted in the center. To determine the final energies,
    one has to take into account the
    different situations noted before, the origin of the excited
    electron (either coming from an $e_g$ or an O-$2p$ state), and
    the fact that putting an extra electron in an already occupied
    $e_g$ level costs the on-site Coulomb repulsion energy $U$.
    When an excited electron comes from an O-site, the charge transfer energy $\Delta$
    has to be added.
  }\label{mang-transitions-fig}
\end{figure}

To facilitate further discussion, we reproduce in
Fig.~\ref{mang-transitions-fig}a the energy diagram proposed for
the manganites (see the figure caption for details).
\cite{millis-review, quijada} The possible transitions, depicted
in Fig.~\ref{mang-transitions-fig}b, are:

\begin{enumerate}
  \item[$-$]  {\it I:} A transition between different $e_g$ levels in the same site.
            This transition is not dipole allowed and, therefore, is
            expected to be weak. \cite{quijada, millis-Fl2polaron} However,
            it has been argued \cite{noh, chen} that due to the strong
            hybridization between the $e_g$ and O-$2p$ bands, and
            the strong local distortion of the Mn-O octahedra,
            these transitions can be allowed.
  \item[$-$]  {\it II, III:} Interatomic $e_g \rightarrow e_g$
            transitions, i.e. from a Mn$^{3+}$ ion to either
            another Mn$^{3+}$ ion or to a Mn$^{4+}$ ion. In both
            cases, the promoted electron can end up being parallel
            or antiparallel to the \tg\ core spin.
  \item[$-$]  {\it IV, V:} Charge transfer transitions, O-$2p \rightarrow e_g$.
\end{enumerate}

Now we address the possible assignments to the peaks recognized by
the DL fit (see Fig. \ref{mang-osc-fig}). Let us start from the
high frequency features. The strong absorption starting at around
2.5 eV is represented by peaks 4 and 5 which, taken together, show
a small temperature dependence and almost no isotope effect. There
is a general consensus that this feature stems from the
charge-transfer transitions ({\it IV} and {\it V}) from the O-$2p$
band to bands of Mn-$d$ character, namely the unoccupied $e_g$
levels. \cite{quijada, noh, okimoto}

The most interesting oscillators are the two low-frequency peaks, 2
($\sim$ 0.5 eV) and 3 ($\sim$ 1.5 eV), which show a strong
temperature dependence (see Fig. \ref{mang-osc-fig}). At high
temperatures, in both samples, only peak 3 is clearly visible and
does not significantly change down to \tc. From this point its
intensity decreases in favor of peak 2. This transfer of spectral
weight is accompanied by a small displacement of peak 2 to low
frequencies and some shift of peak 3 towards high frequencies. This
process stops at around 100 K in S16 and 75 K in S18. Thus, one can
conclude that peak 2 is more favored in the metallic state, while
peak 3 is characteristic of the insulating phase.

Peak 2 most likely corresponds to the processes {\it II} or {\it
III}, where the spin of the final state is parallel to the
$t_{2g}$ core. The fact that it does not form a Drude peak is due
to its polaronic nature, probably enhanced by strains present in
films and polished samples. There has also been some discussion
about whether it transforms from small to large polaron  or
remains small in the whole temperature range (the first has been
suggested from studies in \lcmo\cite{kim-lcmo} and several
compounds of the form A$_{0.7}$R$_{0.3}$MnO$_3$ (A = Nd, La; R =
Sr, Ca)\cite{quijada, millis-Fl2polaron} while the second has been
proposed by studying La$_{2/3}$Ca$_{1/3}$MnO$_3$ thin films). A
careful fit to the corresponding analytical expressions allowed to
recognize, at least in La$_{2/3}$Ca$_{1/3}$MnO$_3$, as always
being a small polaron. \cite{hartinger} The temperature dependence
of its position was found to be much larger than in the case of
La$_{2/3}$Sr$_{1/3}$MnO$_3$ where the polaron seems to be large.
The temperature dependence of the position of peak 2, in both S16
and S18, is more similar to the one found for the small polaron
case.

Peak 3 has been given two interpretations. From studies in various
samples of \lcmo\cite{jung-lcmo}, it has been assigned to a
transition of type {\it I} (see also the theoretical analysis
given in Ref. \onlinecite{chen}). This kind of transition, though
not allowed, can be enhanced by local distortions and strong
hybridization. It was also argued that its spectral weight
decreases with temperature lowering because lattice distortions
also become weaker. \cite{noh} However, a dipole forbidden
transition is unlikely to have such a large spectral weight
possessed by peak 3. To our point of view, a more probable
candidate is an interatomic transition {\it II} or {\it III} where
the final state is antiparallel to the $t_{2g}$ core spin.
\cite{okimoto} This gives a rather small value of $J_H$ of about
0.75 eV. Recent dynamical mean-field calculations \cite{michaelis}
of the critical temperature, spin-wave stiffness, and optical
spectral weight changes provided a two times larger value of
$J_H$, which is nevertheless much closer to 0.75 eV compared to
initial estimates. \cite{chatto} In a homogeneous fully
spin-polarized state the antiparallel process should disappear.
However, we observe, that even at low temperatures the intensity
of peak 3 remains finite. We will see later, that this can
ascribed to the formation of an phase-separated state, which
consists of insulating and conducting domains.

\begin{figure}
  \centerline{\includegraphics[width=86mm,clip=true]{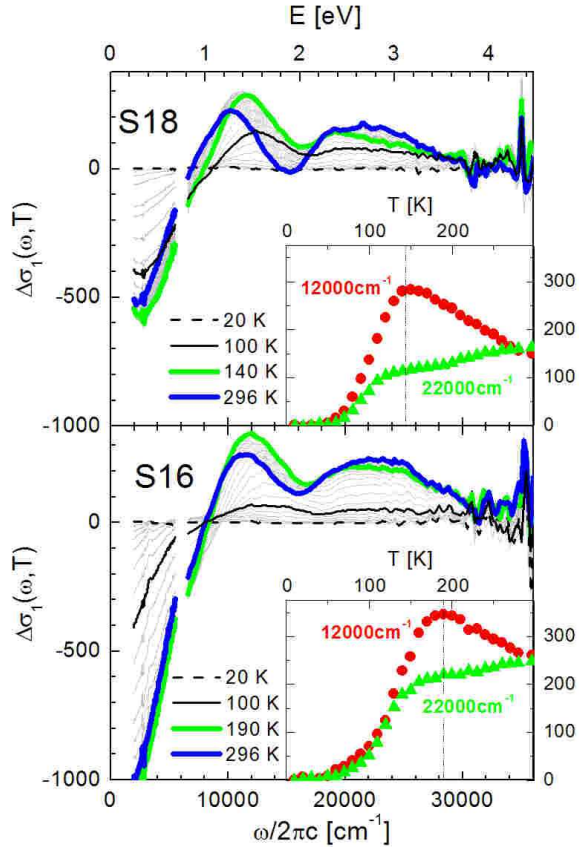}}
  \caption{
    Difference between optical conductivities, $\Delta
    \sigma_1(\omega, T) \equiv \sigma_1(\omega, T) - \sigma_1(\omega, 10 K)$
    for the two samples. There are two clear regions where \sig\
    decreases below \tc. The insets show the temperature
    dependencies at 12000 and 22000 \cm.
  }\label{mang-diff-fig}
\end{figure}

Regarding this assignment of peak 3, we have to mention that in
their study of compounds of the form A$_{0.7}$R$_{0.3}$MnO$_3$ (A =
Nd, La; R = Sr, Ca), Quijada et al.\cite{quijada} associated the
{\it antiparallel} transitions with a feature at 3 eV (giving $J_H
\sim 1.5$ eV) observed on the differential conductivity spectra.
Following the same procedure, we plot the difference between \sig\
at any given temperature and \sig\ at 10 K (see
Fig.~\ref{mang-diff-fig}). As in Ref.~\onlinecite{quijada}, we can
also see that the conductivity in a broad region around 2.7 eV
(22000 \cm) is suppressed below \tc. However, the temperature
dependence of conductivity at 2.7 eV, seen in the insets in
Fig.~\ref{mang-diff-fig}, does not show any feature at \tc\ that
would be natural to expect from transitions which are so sensitive
to the magnetic order. On the other hand, the temperature dependence
at about 1.5 eV (12000 \cm) does show a clear feature at \tc, giving
extra support for our assignment.

Another relevant issue is the importance of the correlation (or
Hubbard-$U$) effects. A large value of $U$ would shift the energy
of the process of type {\it III} or {\it V} to higher energies.
For the $d^4$ configuration, which is the case for the Mn$^{3+}$
atoms, the lower estimate of $U$ is about 1.3 eV if a full
metallic screening is assumed. \cite{marel_sawatzky} In manganites
the screening is expected to be smaller, thus giving an even
larger value of $U$. The analysis of photoemission and X-ray
absorption spectroscopy \cite{park, saitoh} yields $U$ at least
larger than 3 eV. Then we have to conclude that transitions of the
type {\it III} and {\it V} are located at rather high frequencies.
However, in Ref. \onlinecite{michaelis} it has been argued that
the integrated conduction-band spectral weight agrees well with
the band calculations, meaning that the spectral weight transfer
due to the Hubbard-$U$ effects is small.

\subsection{Phase Separation}

\begin{figure}
  \centerline{\includegraphics[width=86mm,clip=true]{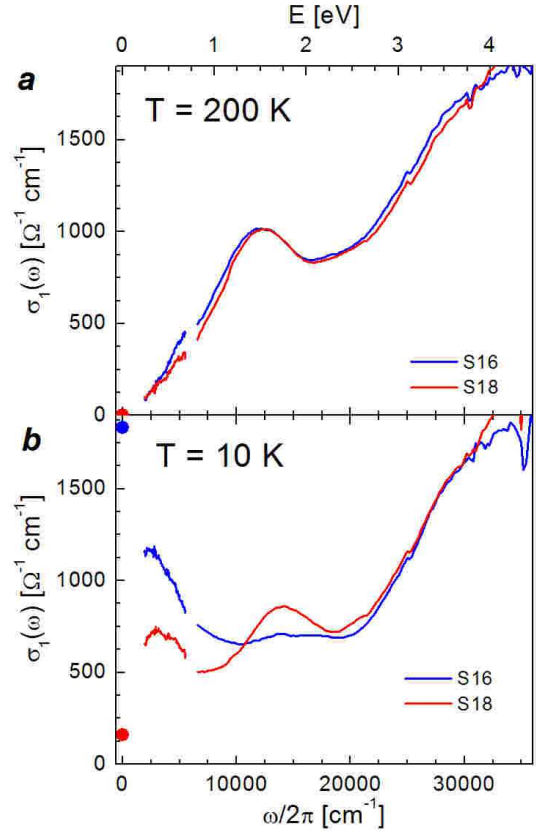}}
  \caption{
    Isotope substitution effect on the optical conductivity in the paramagnetic (200 K) and
    ferromagnetic (10 K) states.
  }\label{mang-s1-fig1}
\end{figure}

Substitution of $^{16}$O with $^{18}$O makes the system less
metallic in the FM state, without affecting it noticeably above \tc.
In accordance, as can be seen in Fig.~\ref{mang-s1-fig1}b, the
conductivity of the sample S18 shows at low temperature a much
larger peak at 1.5 eV (characteristic of the insulating state) and
much smaller peak at 0.5 eV (characteristic of the ferromagnetic
metallic state) compared to the sample S16. What is interesting,
however, is that the effect of the isotope exchange on
finite-frequency conductivity is yet much smaller than the one on
the DC conductivity (the ratio of \sdc's amounts to more than one
order of magnitude for our samples). In view of numerous indications
of the phase separation in manganites (see Ref.~\onlinecite{moreo}
and references therein; see also Ref.~\onlinecite{phase-sep-films}),
it is reasonable to associate the different scale of the isotope
dependence of \sdc\ with the percolation of conducting domains.

\begin{figure}
  \centerline{\includegraphics[width=86mm,clip=true]{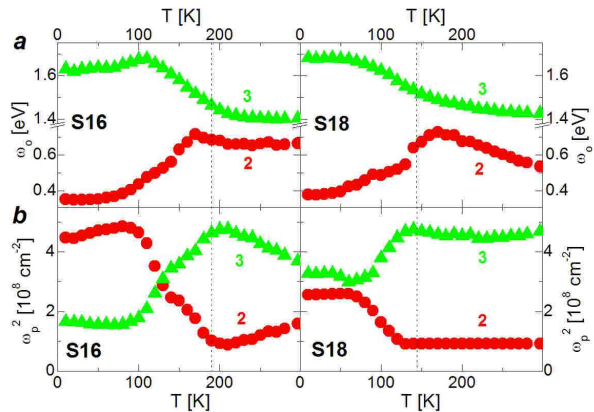}}
  \caption{
    Temperature dependence of the {\it (a)} position, $\omega_o$,
    and {\it (b)} plasma frequency, $\omega_p^2$, of the oscillators number
    2 and 3 (see Fig.~\ref{mang-osc-fig}). Notice also that in sample
    S18 above 110 K, peak 2 almost merges with peak 3. Therefore, above this temperature,
    we have kept $\omega_{p,2}$ constant.
  }\label{mang-par23-fig}
\end{figure}

The idea of phase separation is intimately related with the oxygen
isotope effect itself. In this scenario, \cite{khomskii} at high
temperatures the insulating phase is dominating. When the
temperature is lowered, there is first tendency to charge ordering
(with $T_{CO}
>$ \tc) in the insulating phase. Below \tc, FM metallic droplets
start to form until their relative volume ratio stabilizes. The
temperature variation of the spectral weights of peaks 2 and 3 are
consistent with this interpretation (see Fig.~\ref{mang-par23-fig}).
That is, at low temperatures, the intensity of both peaks tends to
saturate. Moreover, this temperature dependence is similar to the
one seen in the MIR peak of La$_{7/8}$Sr$_{1/8}$MnO$_3$ where it was
also suggested to be caused by phase separation. \cite{jung-lsmo}

The changes in the optical conductivity produced by the isotope
substitution are also consistent with phase separation. In this
picture, the large isotope effect seen in these compounds results
from the fact that the small change induced by the isotope
substitution shifts the relative stability of one phase over the
other, especially close to the phase boundary.
\cite{babushkina-jap83} In the present case, the $^{18}O$
substitution favors the insulating phase. Accordingly, we can see
this effect when comparing the strengths of peaks 2 and 3 between
S16 and S18 at the lowest temperature. In S18 (the {\it less}
metallic of the two), the former oscillator is weaker while the
latter is stronger.

As discussed in Ref.~\onlinecite{babushkina-jap83}, the isotope
substitution changes the effective hopping integral, $t_{eff}$,
which actually determines the relative stability of the different
phases. The change produced by the isotope substitution is small but
can be enhanced if the charge carriers have polaronic nature. This,
as we have seen in the previous section, is the case in the samples
studied here, and may help to explain the large isotope effect.

We also would like to remark that the phase separation scenario
seems to be relevant only to samples with $^{16}$O and somewhat
low content of $^{18}$O.\cite{babushkina-prb62} Studies of the
effect of partial $^{16}$O-$^{18}$O substitution on
(La$_{0.25}$Pr$_{0.75}$)$_{0.7}$Ca$_{0.3}$MnO$_{3}$ ceramic
samples\cite{babushkina-prb62} indicate that the increasing of
$^{18}$O enrichment produced a percolation-like transition to the
insulating at a concentration of around 60\%. Moreover,
magnetic\cite{babushkina-prb62} and neutron
diffraction\cite{balagurov-prb60} measurements indicate that above
60\% there exists only a pure antiferromagnetic insulating phase.
In the same fashion, it is expected that samples of
(La$_{0.5}$Pr$_{0.5}$)$_{0.7}$Ca$_{0.3}$MnO$_{3}$ with $^{18}$O
content larger than 85\% should be antiferromagnetic without phase
separation.

\subsection{Spectral Weight}

\begin{figure}
  \centerline{\includegraphics[width=86mm,clip=true]{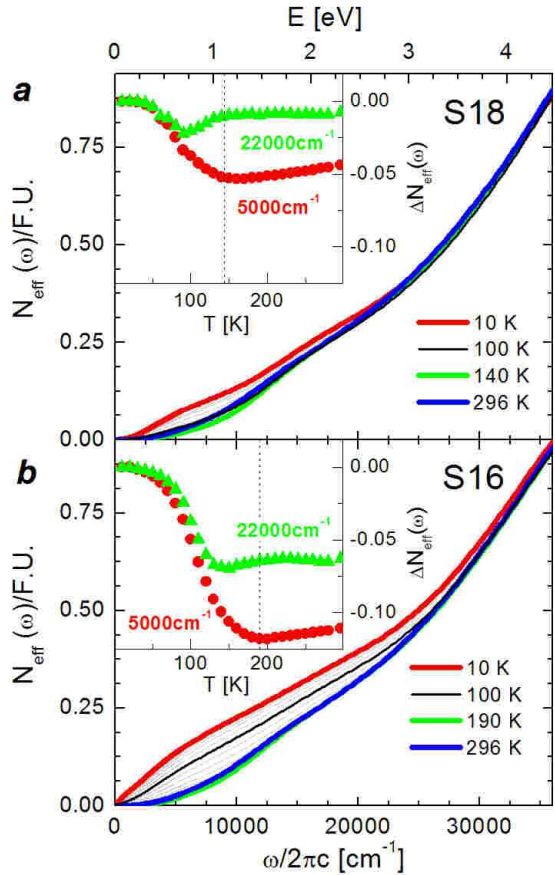}}
  \caption{
    Integrated spectral weight, $N_{eff}$, in the whole measured
    range. The insets show the temperature dependence of
    $\Delta N_{eff} \equiv N_{eff}(T) - N_{eff}(10 \mbox{ K})$
    at two different frequencies.
  }\label{mang-neff-fig}
\end{figure}

Finally, we discuss the integrated spectral weight, which can be
conveniently expressed in terms of the effective number of
carriers:

\begin{equation}\label{mang-neff-eq}
  N_{eff} (\omega) = \frac{2 m V}{\pi e^2} \int_0^\omega \sigma (\omega') d\omega'
\end{equation}

\noindent where $m$ and $e$ are the electron mass and charge,
respectively, and $V$ is the volume of one formula unit. By applying
this formula to the optical conductivity, we obtain the results
presented in Fig. \ref{mang-neff-fig}. The amount and energy scale
of the spectral weight redistribution caused by the metal-insulator
transition can be assessed by the comparison of $N_{eff}(\omega)$
for different temperatures. To facilitate this comparison, in the
insets of the same Figure we have also plotted the temperature
dependence of $\Delta N_{eff} \equiv N_{eff}(T) - N_{eff}(10 \mbox{
K})$ at two different frequencies.

According to the discussion in the previous sections, at
temperatures larger than \tc, the spectral weight is approximately
the same for both samples and shows also small temperature
dependence. Below their respective \tc, the spectral weight at low
frequencies of both samples increase but, at the lowest
temperature, is much smaller in S18 which is also the sample with
smaller \tc. However, we have not found scaling between $N_{eff}$
at 0.5 eV and \tc\ as seen for \lpcmo\ samples with different Pr
concentrations. \cite{kim-lpcmo} Another point to notice is that
the spectral weight in the S18 sample is recovered in a smaller
frequency range than the one containing $^{18}$O.

It is also interesting to compare different compounds where the
distribution of spectral weight has also been studied. In the
manganites studied here, the increase of spectral weight at low
frequencies due to the formation of the metallic state is
recovered at around 4 eV. This is an indication of the width of
the $d$ bands involved in the physical process. In heavy fermion
systems, the spectral weight is recovered in an energy range of
~0.2 eV which indicates the narrowness of the f-bands
participating in the process of forming the heavy fermion coherent
state. \cite{degiorgi} On the other hand, in FeSi the spectral
weight lost in the formation of the Kondo insulating phase is not
recovered below 4 eV. This is also an indication of the width of
the $d$-bands describing its behavior. \cite{schlesinger}

\section{Conclusions}

We have presented the optical properties of two \lpcmoa\ films,
grown on SrTiO$_3$ substrate, each of one contains different oxygen
isotopes. The mismatch between the film and substrate lattice
constants makes the films to be under high strains that change the
their properties  and make them more susceptible to phase separate.
Evidence of this behavior was found in the observed optical response
at low frequencies. In the paramagnetic insulating phase the optical
conductivity of both films is similar and is dominated by a large
peak at 1.5 eV, most probably corresponding to the transition {\it
II} [$e_{g1} (\mbox{Mn}^{3+}) \rightarrow e_g (\mbox{Mn}^{3+})$]
where the moved electron ends up being antiparallel to the \tg\
core. At temperatures below \tc, it looses intensity until it
saturates at around 100 K. In the less metallic sample (the one
containing $^{18}$O), the intensity of this peak at low temperatures
is larger than the more metallic one (the one with $^{16}$O). On the
other hand, the decrease in intensity of this transition is
accompanied by the increase of another peak located at about 0.5 eV.
This latter peak has been identified as the same kind of transition
but where the excited electron ends up being parallel to the spin of
the \tg\ electrons. The intensity of this peak is larger in the more
metallic sample. These observations taken together suggest that
below \tc\ the samples separate in two phases, one ferromagnetic
metallic and one paramagnetic insulating. The relative volume of
these phases is dependent on the isotope content, the metallic phase
being more abundant in the sample containing $^{16}$O.

\section*{Acknowledgements}

We would like to thank D.~I.~Khomskii and A.~J.~Millis for fruitful
discussions, and E.~van~Heumen for their help in performing the
ellipsometry measurements in STO. This investigation was supported
by the Netherlands Foundation for Fundamental Research on Matter
(FOM) with financial aid from the Nederlandse Organisatie voor
Wetenschappelijk Onderzoek (NWO). It was also partially supported by
MaNEP (Switzerland).

\end{document}